\documentclass[aps,prf,preprint]{revtex4-2}
\usepackage{amsmath}
\usepackage{amssymb}
\usepackage{graphicx}
\usepackage{epstopdf, epsfig}
\usepackage{xcolor}
\usepackage{CJK}

\definecolor{mypink}{RGB}{219, 48, 122}
\definecolor{mygreen}{RGB}{0, 128, 0}

\newcommand \red[1]{\textcolor{black}{#1}}
\newcommand \blu[1]{\textcolor{black}{#1}}
\newcommand \ble[1]{\textcolor{black}{#1}}
\newcommand \pink[1]{\textcolor{black}{#1}}
\newcommand \mage[1]{\textcolor{black}{#1}}
\newcommand \new[1]{\textcolor{black}{#1}}
\newcommand \tap[1]{\textcolor{black}{#1}}

\begin{document}
\begin{CJK*}{GB}{}
\title{Stability of variable density rotating flows: Inviscid case and viscous effects in the limit of large Reynolds numbers}

\author{C. Jacques}
\author{B. Di Pierro}
\email{corresponding author: bastien.di-pierro@univ-lyon1.fr}
\affiliation{Univ Lyon, Universit\'e Claude Bernard Lyon I, Ecole Centrale de Lyon,
INSA Lyon, CNRS, Laboratoire de M\'ecanique des Fluides et d'Acoustique, UMR
5509, 43 boulevard du 11 novembre 1918, F-69100, VILLEURBANNE, France}
\author{A. Cadiou}
\affiliation{Univ Lyon, Ecole Centrale de Lyon, INSA Lyon, Universit\'e Claude Bernard
Lyon I, CNRS, Laboratoire de M\'ecanique des Fluides et d'Acoustique, UMR
5509, 36 Avenue Guy de Collongue, F-69134, ECULLY, France}
\author{F. Alizard}
\author{M. Buffat}
\affiliation{Univ Lyon, Universit\'e Claude Bernard Lyon I, Ecole Centrale de Lyon,
INSA Lyon, CNRS, Laboratoire de M\'ecanique des Fluides et d'Acoustique, UMR
5509, 43 boulevard du 11 novembre 1918, F-69100, VILLEURBANNE, France}
\author{L. Le Penven}
\affiliation{Univ Lyon, Ecole Centrale de Lyon, INSA Lyon, Universit\'e Claude Bernard
Lyon I, CNRS, Laboratoire de M\'ecanique des Fluides et d'Acoustique, UMR
5509, 36 Avenue Guy de Collongue, F-69134, ECULLY, France}

\begin{abstract}
Linear stability of solid body rotating \pink{flows} with axisymmetric density variations is addressed analytically.
Considering inviscid disturbances, a non trivial dispersion relation is obtained and it is shown that
the instability is \pink{of} Rayleigh--Taylor type in cylindrical frame. 
The viscous correction is derived, in the limit of large Reynolds numbers and large azimuthal wave numbers,
allowing the determination of the most unstable mode.
Theoretical \pink{predictions} are checked  
\pink{by comparing them to (spectrally accurate)} computed eigenvalues and direct numerical simulations.
\end{abstract}
\maketitle
\end{CJK*}

\section{Introduction}
\pink{Rotating and swirling flows of gas mixtures (\textit{i.e.} variable density flows) are 
encountered in a wide variety of industrial and geophysical flows. 
Rotating injectors and hurricanes are two examples illustrating the 
special case of solid-body rotation \blu{\citep{lumley_1999}}. 
Understanding the dynamics of 
these flows is not only of fundamental interest but is also relevant 
for the development of control and optimisation strategies.}\\
It is well known that solid body rotation of a homogeneous fluid is a marginally
stable configuration of Euler's equations 
\citep{Landau:94,Kelvin:10}, while viscosity \pink{has only a stabilizing influence.}
\ble{Hence, the so-called inertial waves have been} experimentally observed only when the excitation source
is sustained  \citep{McEwan:69}. Considering three dimensional disturbances of a two-dimensional
axisymetric basic configuration,
the flow exhibits an instability if the Rayleigh-discriminant 
$\Phi = 2V(dV/dr + V/r)/r$ is negative somewhere in the flow \citep{Leibovich:83}
 ($r$ being the distance from rotation center and $V(r)$ the azimuthal
velocity).\\ 
If density variations within the fluid are considered,
an equivalent
to the Br\"unt-Va\"issalla frequency $G$ is usually defined by
$G^2 = -(\Omega^2 r/\rho) d\rho/dr$ ($\rho$ being the density and $\Omega=V/r$).
Leibovith \cite{Leibovich:69} and Howard \cite{Howard:73} have shown independently
that the flow is stable to axisymmetric disturbances if $G^2<\Phi$, 
for both incompressible and compressible flows. For
rotating flows with axial jet, Leibovitch and Stewartson \cite{Leibovich:83}
have been able to theoretically derive the discrete spectrum of the instability near the point
where the Doppler frequency $kW+m\Omega - \omega$ admits an extremum ($\omega,~k,~m,~W$ being  the
pulsation, axial wavenumber, azimuthal wavenumber and axial velocity, respectively). 
Subsequently, Eckhoff \cite{Eckhoff:84}
derived a sufficient condition for instability using a WKB expansion. 
Those results have been extended by Leblanc and Le Duc \cite{leblanc:2005},
who established a link between the eigenfrequencies of 
\blu{\cite{Eckhoff:84} and \cite{Leibovich:83}}.
Di Pierro and Abid \cite{Dipierro:2010} have proposed an asymptotic expression for the
eigenfrequency associated with rotating flow with axial jet, assuming large wavenumbers.\\
Along with that, Gans \cite{Gans:75} has shown that the flow is
unstable with respect to any non axisymmetric two-dimensional disturbances if
$G^2$ is slightly positive. Additionally, considering an equivalent configuration, Sipp \textit{et al.} \cite{Sipp:2005} have 
shown through numerical experiments that $G^2>0$ is
a necessary condition for such an instability. The authors \red{argued} that the underlying mechanism
was associated with a Rayleigh--Taylor instability. \\
\tap{More recently, Scase and Hill \cite{scase:2018} investigated both theoretically and numerically the effect
of rotation on confined two liquid layers, with different density and viscosity, that form concentric cylinders.
Their axis of rotation is confused with central axes of these cylinders. The authors have also discussed the
effect of rotation of both miscible and immiscible fluids.
For all flow cases, the equilibrium state is associated with the two-layers initially separated by a sharp interface.
An Orr-Sommerfeld like equation is derived that successfully reproduces 
Direct Numerical Simulations as far as the linear approximation is still 
valid. Especially, they have shown that the system may be unstable when the 
inner fluid is denser than the outer one driven by the
centrifugal force. The instability takes the form of mushroom like 
perturbations in the nonlinear regime as observed in Rayleigh--Taylor 
configuration. Scase and Sengupta \cite{scase:2021} extended these results to a three dimensional configuration.
They showed that the disturbances growth is essentially driven by geometrical parameters in the absence of
surface tension: domain aspect ratio and initial position of the interface.
While the above findings provide a good understanding of the 
amplification of Rayleigh--Taylor like instability
in a centrifugally driven configuration, it does not discuss the case of 
initial gas mixtures and spatially extended configuration in the radial 
direction. 
Finally, an equivalent configuration of a cylindrical biphasic flow
with uniform radial acceleration has been studied by Zeng \textit{et al.} \cite{Zeng:2020}.
They observed both two and three-dimensional disturbances and particularly, they
showed that the growth rate of the two-dimensional mode varies as $\nu^{1/3}$ ($\nu$ the
kinematic viscosity), while
the three dimensional regime is very dependent of the viscosity ratio.}\\
\mage{
The purpose of this paper is then to investigate certain aspects of Rayleigh--Taylor instability in
rotating flows and to extend previously mentionned studies. The studied configuration is a steady, solid-body rotation and incompressible flow
in the presence of a radial smooth density gradient. Two results are obtained : i)
an analytical dispersion relation without any asymptotic consideration in the inviscid case
and  ii) a viscous correction in the limit of large Reynolds numbers. 
The present study allows an extension of some conclusions in \cite{Dipierro:2010}, which are limited
to a large wavenumber and inviscid analysis. This paper also extends the study of \cite{Sipp:2005} which numerically deals with
the stability of inviscid Gaussian vortex with heavy cores. The present study corresponds to the case
of non axial flow $W(r)=0$ of \cite{Dipierro:2010} and small density core length $b\rightarrow 0$
of \cite{Sipp:2005}. In particular, it is shown that a sufficient condition for instability is $G^2>0$ 
somewhere in the flow for any density ratio, and that the viscous correction has a purely stabilizing effect,
allowing the determination of the most unstable mode. A very good agreement between the analytical
 dispersion relation and (linear/nonlinear) numerical simulations is obtained.}\\
\blu{The paper is organized as follows. In section \ref{sec::problem}, the
problem formulation is stated with mathematical and numerical considerations.
Section \ref{sec::inviscid}  establishes the inviscid
dispersion relation, and section \ref{sec::viscous} deals with the viscous
correction. Both are validated by comparison with computed eigenvalues in these two sections. Finally, those results are
compared with direct numerical simulations in section \ref{sec::dns}, and
the last section provides some physical discussion and concluding remarks.}

\section{Problem formulation}
\label{sec::problem}

\subsection{Governing equations}
\begin{figure}
\centering
\includegraphics[width=0.32\linewidth]{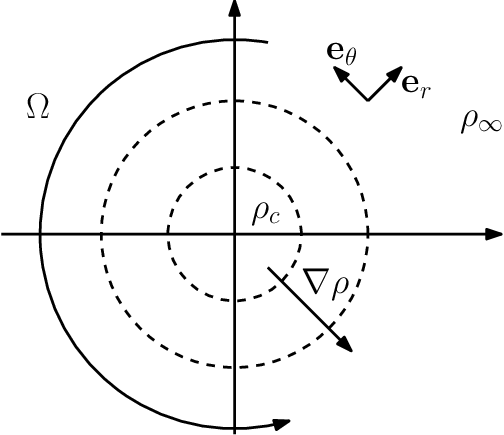}
\caption{Sketch of the basic flow.}
\label{fig:sketch}
\end{figure}

The incompressible, variable-density Navier-Stokes equations are considered, 
neglecting molecular diffusion \citep{guillen::07}:
\begin{eqnarray}
\frac{\partial \mathbf{u}}{\partial t} &=& -(\mathbf{u}\cdot\nabla) \mathbf{u} - \frac{1}{\rho}\nabla P + \frac{\mu}{\rho} \Delta \mathbf{u},\label{eq::NSu} \\
\frac{\partial \rho}{\partial t} &=& -\mathbf{u}\cdot\nabla \rho, \label{eq::NSrho} \\
\nabla \cdot \mathbf{u} &=&0, \label{eq::divu}
\end{eqnarray}
with $\mathbf{u}$ the vector velocity field, $\rho$ the density and 
$P$ the pressure. \red{The dynamic viscosity $\mu$ is assumed to be constant, 
which is representative of gazeous or liquid mixtures,} \pink{ and the surface tension is neglected}.  We investigate a steady solid-body rotation 
base flow with rotation rate $\Omega$ and an axisymmetric density profile, as
sketched in figure \ref{fig:sketch}. 
\new{By introducing $R$, the density core length, as the
length scale and $\Omega^{-1}$ as the time scale, one can define the Reynolds number
$Re = \rho_\infty \Omega R^2/\mu$ where $\rho_\infty$ is the density far from the 
rotating center. }
\pink{The velocity, 
pressure and density fields of the base flow are expressed in polar coordinates $(r,\theta)$ as 
$\mathbf{U_0}(r) = \Omega r \mathbf{e}_\theta$, $\red{P_0(r)}$ and $\rho_0(r)$
, respectively.} 
We consider two-dimensional  disturbances of the form :
\begin{equation}
(u_r(\mathbf{x}, t),~u_\theta(\mathbf{x}, t),~\rho(\mathbf{x}, t), ~\red{P(\mathbf{x}, t)})^T = (u(r),~v(r),~\rho(r), ~\red{p(r)})^T \exp(i(m\theta - \omega t)),
\end{equation}
with $m$ the azimuthal wavenumber and $\omega$ the unknown complex pulsation.
Lineari\ble{zation of} equations (\red{\ref{eq::NSu}, \ref{eq::NSrho}, \ref{eq::divu}}) leads to the following eigensystem :
\begin{eqnarray}
 i\left (m\Omega - \omega +i\frac{\mathcal{L}}{Re\rho_0}\right )u - 2\left (\Omega - \frac{im}{Re\rho_0 r^2} \right) v + \frac{1}{\rho_0}\frac{d P}{d r} - \frac{\Omega^2 r}{\rho_0} \rho = 0, \nonumber\\
 2\left (\Omega - \frac{im}{Re\rho_0 r^2} \right)u + i\left (m\Omega - \omega +i\frac{\mathcal{L}}{Re\rho_0}\right ) v + \frac{im}{r \rho_0} P = 0, \nonumber\\
i\left (m\Omega - \omega \right ) \rho + u \frac{d\rho_0}{dr} = 0,\nonumber\\
\frac{1}{r}\frac{d ru}{dr} + \frac{im}{r} v =0,
\label{eq::NSlin}
\end{eqnarray}

with $\displaystyle{\mathcal{L} = \frac{d^2}{dr^2} + \frac{1}{r}\frac{d}{dr} - \frac{m^2+1}{r^2}}$.

\subsection{Numerical resolution}
The \blu{eigenvalues and eigenfunctions $(u,v,\rho,p)$ are computed with 
spectral accuracy by using Chebyshev polynomial expansions.}
\pink{The infinite domain $r\in[-\infty,\infty]$ is mapped to the interval 
$\gamma \in[-1,1]$ containing the Gauss-Lobato points ($\gamma_i=\cos(i\pi/(N-1)),~i=0..N-1$) using the 
\ble{transformation} $r=tan(\pi\gamma/2)$.}
\new{This transformation avoid the expression of a boundary condition at $r=0$, which
depend on the disturbance symmetry. The number of modes $N$ is even such that the singular point $r=0$
is avoided.}
\pink{In the following , the radial derivative operators of order $n$ 
will be denoted by $D_r^n$ \citep{peyret2013}}.\\
\new{Since the equation associated with the pressure is independant of $\omega$, the
system \ref{eq::NSlin} is numerically ill-conditioned. This leads to the computation
of non physical oscillating modes \citep{peyret2013}}.
To avoid \pink{the occurence} of these spurious modes \pink{when solving the 
eigensystem}, the pressure is eliminated
from (\ref{eq::NSlin}) by taking the divergence of the momentum equations
and \blu{using} the inverse of an elliptic operator:
\begin{equation}
\mathbb{D} =\left [ \frac{m^2}{r^2\rho_0} - \left( D_r + \frac{1}{r} \right ) \frac{1}{\rho_0} D_r \right]^{-1}.
\end{equation}
Now, the eigenvalue problem \blu{can be recast as:}
\begin{equation}
\mathcal{A}
\begin{pmatrix}
u\\
v\\
\rho
\end{pmatrix}
=
i \omega
\begin{pmatrix}
u\\
v\\
\rho
\end{pmatrix},
\label{eq::full}
\end{equation}
with
\[\mathcal{A}=
\begin{bmatrix}
im\Omega - \frac{1}{Re\rho_0}\mathbb{L} +\frac{1}{\rho_0}D_r\mathbb{D}C_u & -2\left (\Omega - \frac{im}{Re\rho_0 r^2} \right ) + \frac{1}{\rho_0}D_r\mathbb{D}C_v & \frac{-\Omega^2r}{\rho_0} + \frac{1}{\rho_0}D_r\mathbb{D}C_\rho\\
2\left (\Omega - \frac{im}{Re\rho_0 r^2} \right ) + \frac{im}{\rho_0r}\mathbb{D}C_u & im\Omega - \frac{1}{Re\rho_0}\mathbb{L} +\frac{im}{\rho_0r}\mathbb{D}C_v & \frac{im}{\rho_0r}\mathbb{D}C_\rho \\
\frac{d\rho_0}{dr} & 0 & im\Omega
\end{bmatrix},
\] 

\[
\left \{
\begin{matrix}
C_u &=& \left ( D_r + \frac{1}{r} \right ) \left (im\Omega - \frac{1}{Re\rho_0}\mathbb{L} \right ) + 2 \frac{im}{r}\left(\Omega - \frac{im}{r^2Re\rho_0} \right ) \\
C_v &=& -2\left ( D_r + \frac{1}{r} \right ) \left(\Omega - \frac{im}{r^2Re\rho_0} \right ) + \frac{im}{r} \left (im\Omega - \frac{1}{Re\rho_0}\mathbb{L} \right )\\
C_\rho &=& -\left ( D_r + \frac{1}{r} \right ) \frac{\Omega^2 r}{\rho_0}
\end{matrix}
\right .
\]
 and
$\displaystyle{\mathbb{L} = D_r^2 + (1/r)D_r - (m^2+1)/r^2}$. 
In addition to eliminating spurious modes, \red{elimination of the pressure improves operator conditionning}
and accelerates the computation of eigenvalues. The last
eigensystem (\ref{eq::full}) is solved using LAPACK routines.

For numerical validation, the following basic density profile is used :
\begin{equation}
\red{
\rho_0(r) = 1+\frac{s-1}{2}\left ( 1 - \tanh \left (\frac{r-1}{e} \right ) \right )}, \label{eq::rhoBase}
\end{equation}
which mimics heavy (or light) vortex core with smooth density variations; $e$ being
the variation length of density profile \new{and $s$ being the density ratio between the
rotating center and the far field.}

\section{Inviscid instability}
\label{sec::inviscid}
In the limit of an infinite Reynolds number,
 the system (\ref{eq::NSlin}) can be rewritten as a single equation of the
variable $\phi = ru$, as:
\begin{equation}
\frac{d^2 \phi}{dr^2} + \frac{1}{\rho_0r}\frac{d\rho_0 r}{dr}\frac{d\phi}{dr} - \left [ \frac{2m\Omega}{r\rho_0\Sigma} \frac{d\rho_0}{dr} + \frac{m^2}{r^2} \left ( 1+ \frac{G^2}{\Sigma^2} \right ) \right ]\phi=0, \label{eq::phi}
\end{equation}
with:
\begin{equation}
G^2 = -\frac{\Omega^2 r}{\rho_0}\frac{d \rho_0}{dr}, ~~~~ \Sigma = m\Omega-\omega. \label{eq::defG2}
\end{equation}

Introducing $\psi = \sqrt{r \rho_0}\phi$, equation (\ref{eq::phi}) becomes:
\begin{equation}
\frac{d^2 \psi}{dr^2} +\left [\zeta -  \frac{2m\Omega}{r\rho_0\Sigma} \frac{d\rho_0}{dr} - \frac{m^2}{r^2} \left ( 1+ \frac{G^2}{\Sigma^2} \right ) \right ]\psi = 0, \label{eq::psi}
\end{equation}
with:
\begin{equation}
\zeta = \left [ \frac{1}{2\rho_0 r} \frac{d \rho_0 r}{dr} \right ]^2 - \frac{1}{2\rho_0 r}\frac{d^2 \rho_0 r}{dr^2} .
\end{equation}
From (\ref{eq::psi}), \blu{we assume} that the unstable dynamics occurs at the location
$r^*$, such that $G^2(r^*)$ is \ble{a local}  extremum. 
Introducing, $\tilde{r} = r-r^*$ and $\tilde{\psi}(\tilde{r}) = \psi(r)$, a second order
Taylor expansion of (\ref{eq::psi}) gives:
\begin{equation}
\frac{d^2 \tilde{\psi}}{d\tilde{r}^2} + \Theta(\tilde{r}) \tilde{\psi} = 0 \label{eq::psi_dev}
\end{equation}
with:
\begin{eqnarray}
\Theta(\tilde{r}) &=& \zeta(r^*) - \frac{\chi(r^*)}{r^{*2}} + \left ( \frac{d\zeta}{dr}(r^*) + \frac{2\chi(r^*)}{r^{*3}} \right ) \tilde{r} + \left (\frac{d^2\zeta}{dr^2}(r^*) - \frac{1}{r^{*2}}\frac{d^2\chi}{dr^2}(r^*) - \frac{6\chi(r^*)}{r^{*4}} \right ) \frac{\tilde{r}^2}{2}, \nonumber \\
\chi(r) &=& 2m\frac{G^2}{\Omega\Sigma} + m^2\left (1+\frac{G^2}{\Sigma^2} \right ).
\end{eqnarray}
Equation (\ref{eq::psi_dev}) is a parabolic cylinder equation. Assuming that
$\psi$ vanishes far from $r^*$, equation (\ref{eq::psi_dev})
\pink{admits solutions in the form of Weber-Hermite polynomials}
$\tilde{\psi}(\tilde{r}) = H_n(\alpha\tilde{r}+\beta)$ with:
\begin{equation}
\alpha = \sqrt{2}\left (3\frac{\chi(r^*)}{r^{*4}} + \frac{1}{2r^{*2}}\frac{d^2\chi}{dr^2}(r^*) - \frac{1}{2}\frac{d^2\zeta}{dr^2}(r^*) \right  ) ^{1/4}, ~~\beta = \frac{2}{\alpha^3} \left ( \frac{d\zeta}{dr}(r^*) + \frac{2\chi(r^*)}{r^{*3}} \right ),
\end{equation}
if and only if:
\begin{equation}
-\frac{1}{\alpha^2}\left (\frac{\chi(r^*)}{r^{*2}} - \zeta(r^*)  + \frac{ \left ( \frac{d\zeta}{dr}(r^*) + \frac{2\chi(r^*)}{r^{*3}} \right )^2}{2\frac{d^2\zeta}{dr^2}(r^*) - 2\frac{1}{r^{*2}}\frac{d^2\chi}{dr^2}(r^*) - \frac{12\chi(r^*)}{r^{*3}} } \right ) = n+\frac{1}{2},\label{eq::relDispInv}
\end{equation}
where $n$ is an integer. 
\red{This inviscid dispersion relation \ble{corresponds to the case studied numerically}
in \cite{Sipp:2005} for small density distribution widths
(defined as ``$b$'' in their paper). 
Note that it is not based on the assumption of a weakly varying base flow 
\citep{Dipierro:2010} or large wavenumbers.} 
In the limit of large azimuthal wavenumbers, the following growth
rate is obtained by expansion of (\ref{eq::relDispInv}) with respect to $m$:
\begin{equation}
\omega_n = m\Omega + \frac{G^2}{m\Omega} +i \left (\sqrt{G^2(r^*)} + \frac{n+\frac{1}{2}}{m} \sqrt{-\frac{r^{*2}}{2}\frac{d^2G^2}{dr^2}(r^*)} \right ). \label{eq::relDispInvAsymp}
\end{equation}
Here, the expression (\ref{eq::relDispInvAsymp}) is identical to the one obtained in
\cite{Dipierro:2010}, except for the correction $G^2/m\Omega$ to the real part.
Equation (\ref{eq::relDispInv}) has no explicit solution, and is numerically solved using 
the Muller's root finding algorithm \blu{\citep{lang1994}}. 

\begin{figure}
\centering
\includegraphics[width=0.5\linewidth]{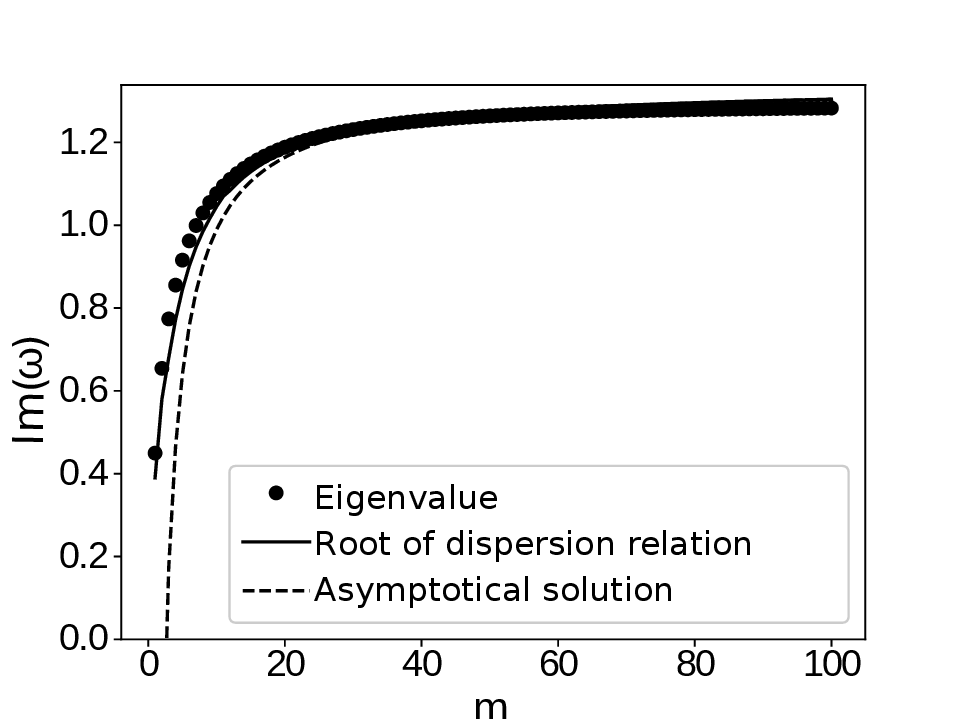}
\caption{Inviscid dispersion relation $Im(\omega)$ versus azimuthal wavenumber $m$ for
density ratio $s=2$ and mass thickness $e=0.2$ (Dotted: eigenvalue of system (\ref{eq::full}), line:
root of dispersion relation (\ref{eq::relDispInv}), dashed: Asymptotical solution (\ref{eq::relDispInvAsymp})). }
\label{fig::relDispInv}
\end{figure}

\pink{Figure \ref{fig::relDispInv} compares the dispersion relation obtained 
using the full system (\ref{eq::full}) to the solution of (\ref{eq::relDispInv}) and the 
asymptotic expression (\ref{eq::relDispInvAsymp}).}
A very good agreement is found: the solutions of system (\ref{eq::full}) and equation (\ref{eq::relDispInv})
are almost undistinguishable, and
the relative error is less than $10\%$ if $m\geq6$ when comparing the solution of (\ref{eq::full})
and (\ref{eq::relDispInvAsymp}).

\section{Asymptotic viscous instability}
\label{sec::viscous}
\begin{figure}
\centering
\includegraphics[width=0.45\linewidth]{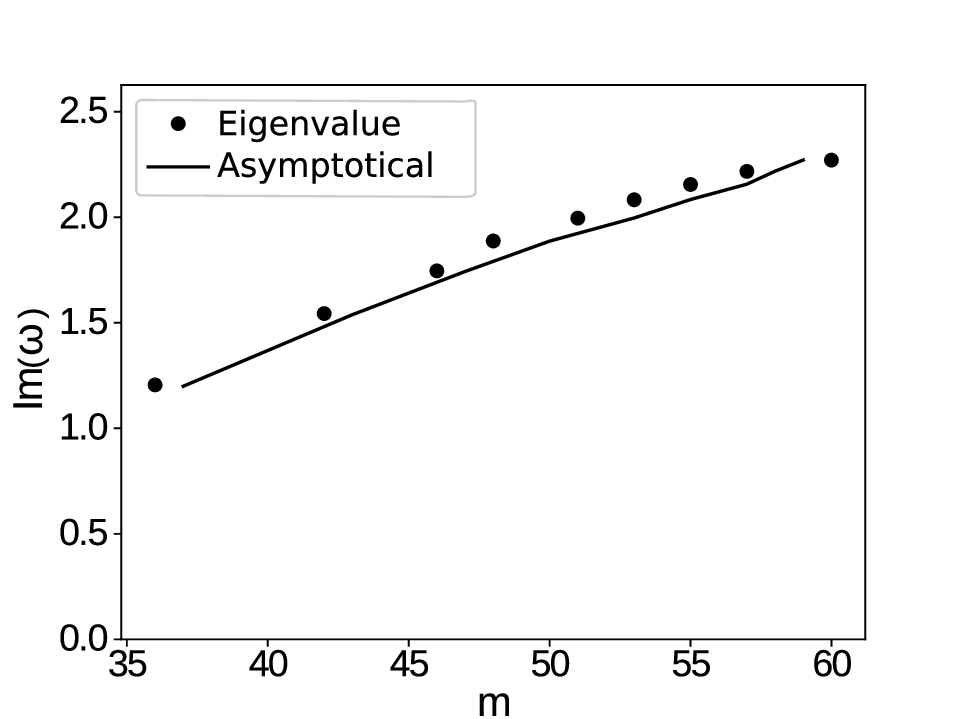}
\includegraphics[width=0.45\linewidth]{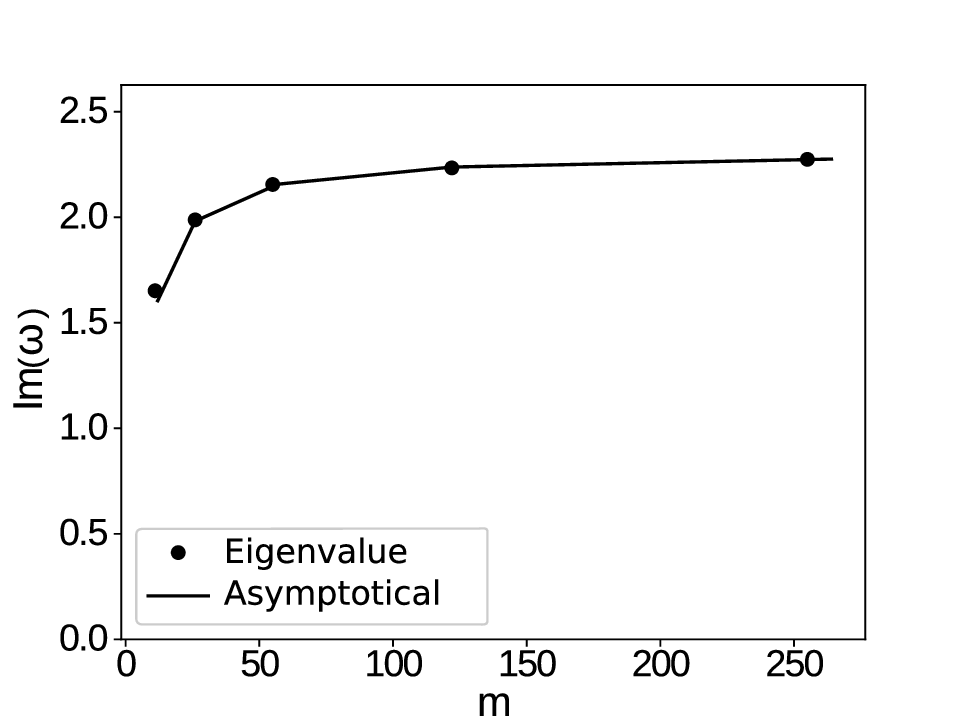}
\caption{Numerical and asymptotical growth rate $Im(\omega_{n, max})$ versus $m_{max}$ for the most unstable mode for $e=0.2$.
Left: $Re=10^4$ and $\red{s}=2,~3,~4,~5,~6,~7,~8,~9,~10$. Right: $s=8$ and $Re=10^2,~10^3,~10^4,~10^5,~10^6$ (Dotted: eigenvalue
of system (\ref{eq::full}), line: Asymptotical solution (\ref{eq::Omega_most_unstable}-\ref{eq::M_most_unstable})).}
\label{fig::comp_mmax_sigmax}
\end{figure}

The viscous case is considered here in the limit of large Reynolds numbers
as well as large wave numbers, such that \red{$1\ll m^2 \ll Re$. Hence, 
$m^2Re^{-1}$ is treated as a small parameter}. In this
limit, the viscous operator behaves to leading order as
\begin{equation}
\frac{1}{Re}\mathcal{L} \approx \frac{-m^2}{r^2Re}.
\end{equation} 
The differential system (\ref{eq::NSlin}) can then be rewritten 
\ble{as } 
a second order differential equation:
\begin{eqnarray}
\frac{d^2 \phi}{dr^2} &+& \left[\frac{1}{\rho_0r}\frac{d\rho_0 r}{dr}+\frac{1}{\Sigma + i\mathcal{N}}\frac{d\mathcal{N}}{dr} \right]\frac{d\phi}{dr} \nonumber \\
 &-& \left [ \frac{2m(\Omega - i\mathcal{M})}{r\rho_0(\Sigma+i\mathcal{N})} \frac{d\rho_0}{dr} + \frac{m^2}{r^2} \left ( 1+ \frac{G^2}{\Sigma(\Sigma+i\mathcal{N})} \right ) - \frac{2im}{r}\frac{1}{\Sigma+i\mathcal{N}} \frac{d\mathcal{M}}{dr} \right ]\phi=0,~~ \label{eq::phivis}
\end{eqnarray}
with:
\begin{equation}
\mathcal{N} = \frac{-m^2}{r^2Re}, ~~~~ \mathcal{M} = \frac{m}{r^2Re}.
\end{equation}
\pink{Proceeding in the same way as in previous section}, 
one gets to leading order in $m$ and $m^2/Re$:
\begin{eqnarray}
\frac{d^2 \psi}{dr^2}-\left [ \frac{2mG^2}{r^2\Omega (\Sigma+i\mathcal{N})}+ \frac{m^2}{r^2} \left ( 1+ \frac{G^2}{\Sigma(\Sigma+i\mathcal{N})} \right )\right  ] \psi=0,~~ \label{eq::phivisAsymp}
\end{eqnarray} 
with $\psi = \sqrt{\rho_0r(\Sigma +i\mathcal{N})}\phi$.
\pink{Here again, a parabolic cylinder equation is found, whose solutions are Weber-Hermite polynomials associated with the following dispersion relation:}
\begin{equation}
\omega_n = m\Omega + \frac{G^2(r^*)}{\Omega m} + i\left (\sqrt{G^2(r^*)} - \frac{n+1/2}{m}\sqrt{-\frac{r^{*2}}{2}\frac{d^2G^2}{dr^2}(r^*)} - \frac{m^2}{2Re\rho_0(r^*)r^{*2}} \right ).
\label{eq::asympVisq}
\end{equation}
\begin{figure}
\centering
\includegraphics[width=0.5\linewidth]{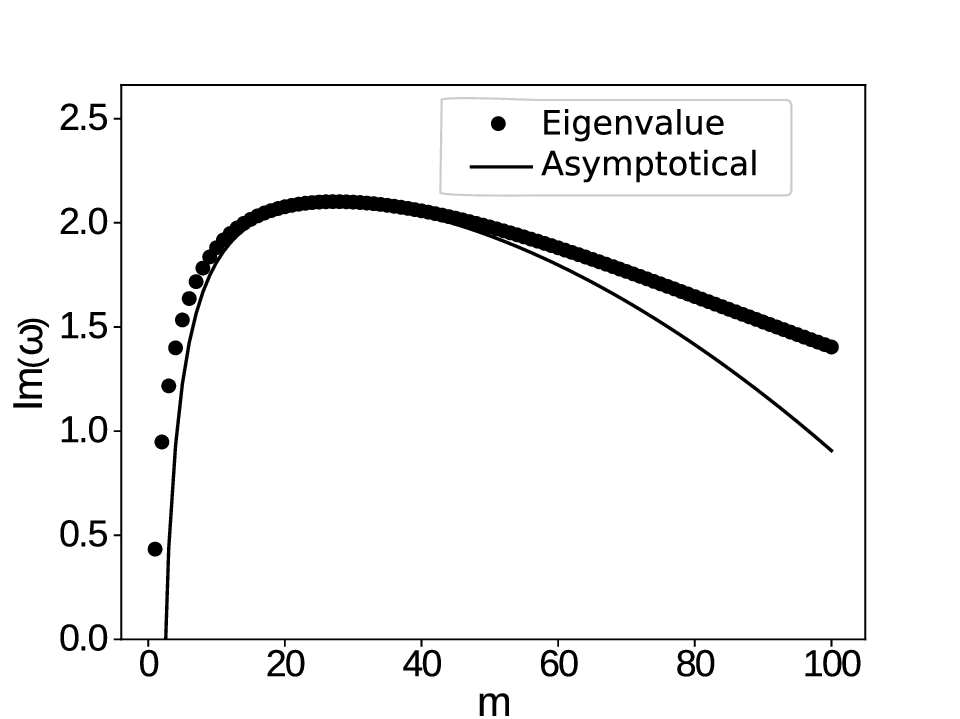}
\caption{Growth rate $Im(\omega)$ versus azimuthal wavenumber $m$ for $s=10$, $e=0.2$ and $Re=1000$ (Dotted: eigenvalue
of system (\ref{eq::full}), line: Asymptotical solution (\ref{eq::asympVisq})).}
\label{fig::validAsympVisq}
\end{figure}
Figure \ref{fig::validAsympVisq} shows the imaginary part of the eigenfrequencies obtained numerically 
in comparison with the asymptotic solution (the real parts being undistinguishable). 
It appears that equation (\ref{eq::asympVisq}) is a good approximation when $12\leq m \leq 50$ (the error is less
than $2.5 \%$); \red{ \textit{i.e.} in the limit of large $m$ and $m^2Re^{-1}\lesssim O(1)$}. Moreover, 
one can see that the maximum of the dispersion \pink{curve (see figure \ref{fig::validAsympVisq})}  is correctly represented by this last
approximation.
This allows \pink{the identification of} the most unstable mode as well as the corresponding azimuthal wavenumber:
\begin{eqnarray}
\omega_{n, max} &=& m\Omega + \frac{G^2(r^*)}{\Omega m} + i\sqrt{G^2} - \frac{3i}{2}\frac{\left(\left (n+\frac{1}{2}\right){\sqrt{-\frac{r^{*2}}{2}\frac{d^2G^2}{dr^2}(r^*)}} \right )^\frac{2}{3} }{(r^{*2}\rho_0(r^*)Re)^\frac{1}{3}}  \label{eq::Omega_most_unstable},\\
m_{max} &=& \left (\left (n+\frac{1}{2} \right ) Re\rho_0(r^*)r^{*2} \right )^{\frac{1}{3}} \left (-\frac{r^{*2}}{2}\frac{d^2 G^2 (r^*)}{dr^2} \right )^{\frac{1}{6}}. \label{eq::M_most_unstable}
\end{eqnarray}
A comparison between these approximations and numerical solutions is shown in figure \ref{fig::comp_mmax_sigmax}. 
Very good agreement is found for the
most amplified mode $\omega_{n, max}$ as well as for the selected azimuthal mode $m_{n, max}$.

\section{Nonlinear instability}
\label{sec::dns}
\begin{figure}
\centering
\includegraphics[width=0.45\linewidth]{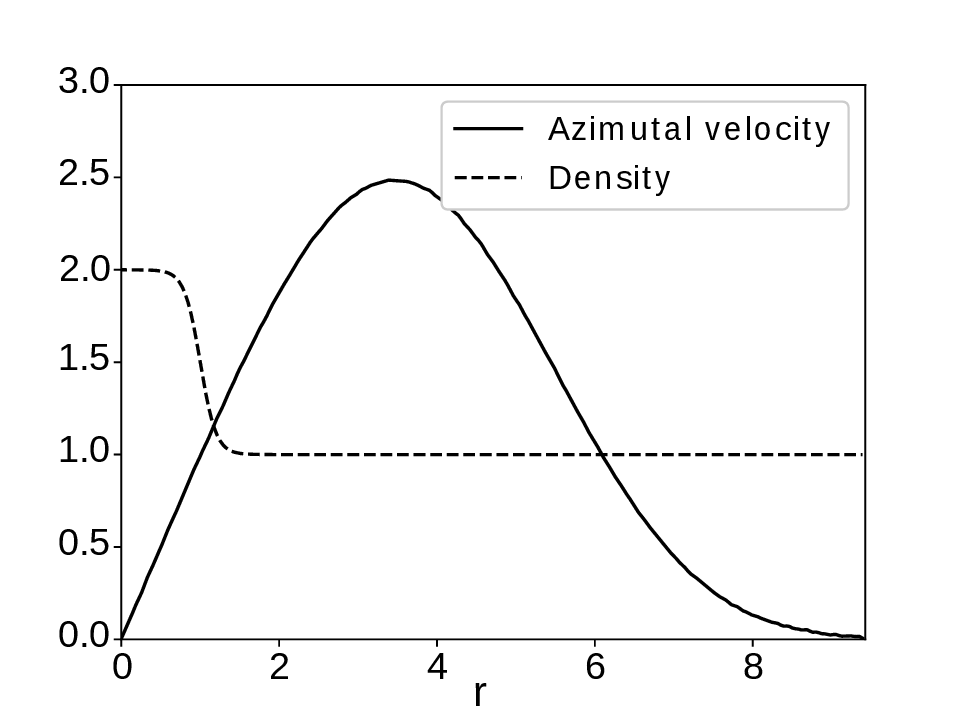}
\includegraphics[width=0.45\linewidth]{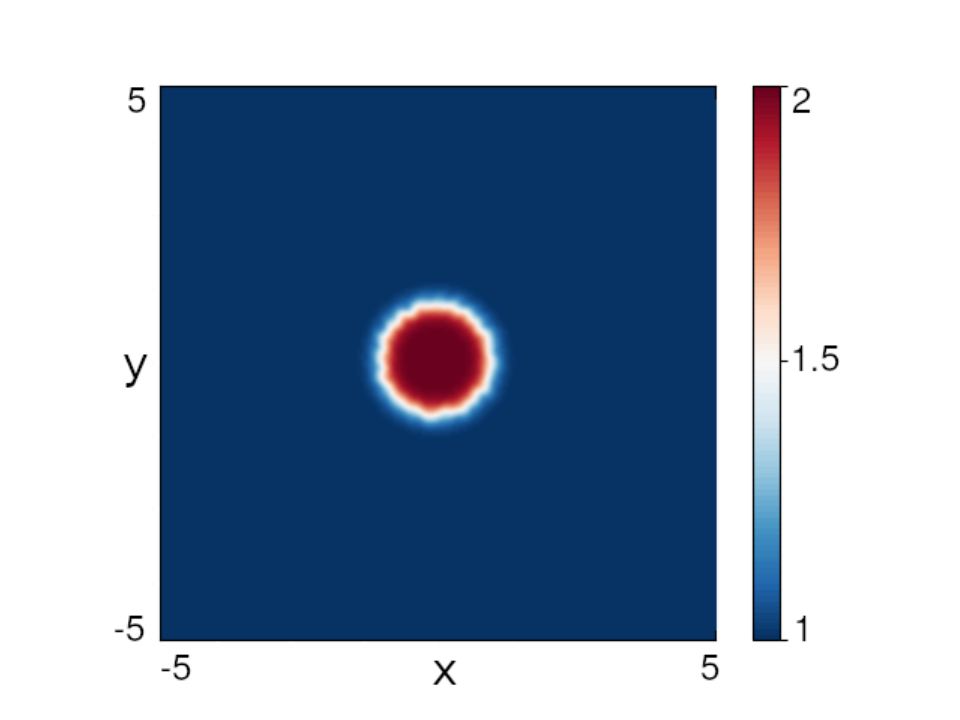}
\includegraphics[width=0.45\linewidth]{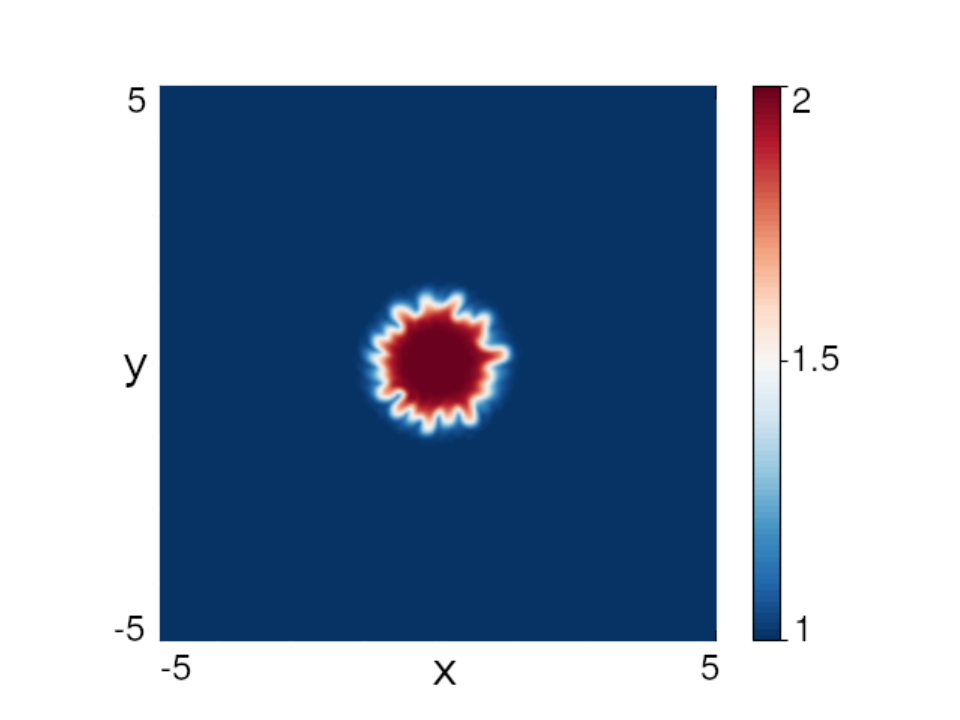}
\includegraphics[width=0.45\linewidth]{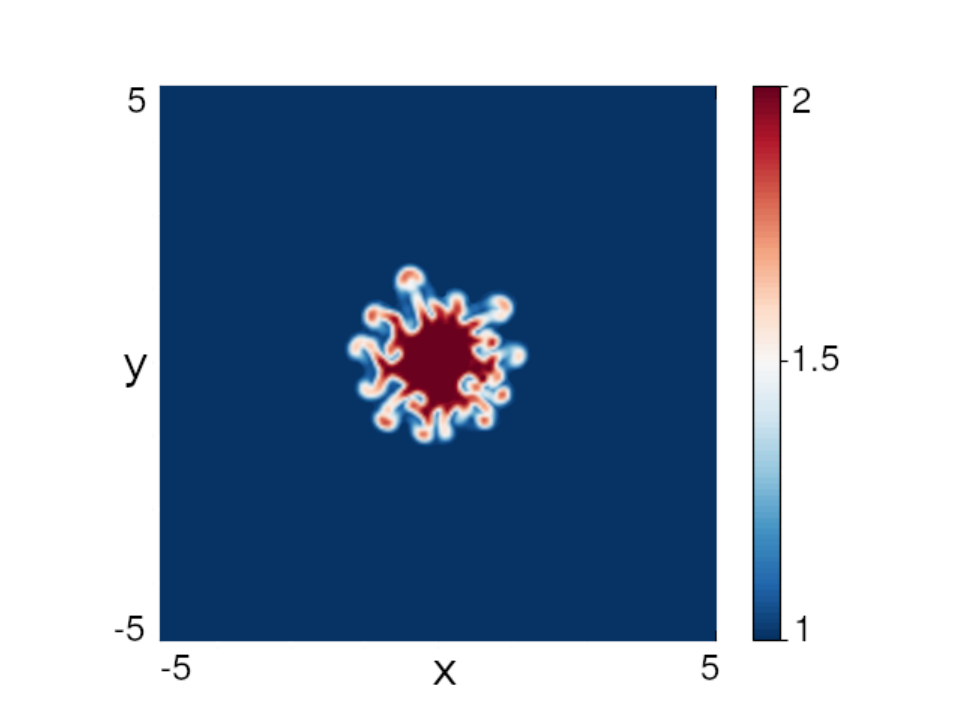}
\caption{Direct numerical simulation of the variable density rotating flow with $s=2$ and $Re=1000$. Initial radial distribution (top left), 
Iso-contours of density field at time 
$t\approx3.7$ (top right),$t\approx5.5$ (bottom left) and $t\approx7.4$ (bottom right). $x$ and $y$ are here the Cartesian coordinates.}
\label{fig::DNS_TI}
\end{figure}

To analyse the instability dynamics \pink{towards} the non linear stage, the 
variable density Navier--Stokes equations (\red{\ref{eq::NSu}, \ref{eq::NSrho}, \ref{eq::divu}})
are \pink{solved numerically} using the method presented in \cite{dipierro:13}. Here is a brief review. The spatial
discretisation \pink{is based on a two-dimensional Fourier expansion,} 
\red{with periodic boundary conditions}. 
Time integration is performed using
a second order \red{Runge-Kutta} scheme, while the pressure is computed 
using a 
\red{fixed point method to ensure the incompressiblity constraint}. 
The time step is choosen as $10^{-2}$ \blu{in order to satisfy stability
criterion,} and the computational domain \pink{is a square of size}
$6\pi \times 6 \pi$ \red{discretized with $512\times512$ grid points}. 
The initial density
profile (\ref{eq::rhoBase}) is used. The velocity initial condition is a Lamb--Oseen \red{like} vortex:
\red{
$\Delta(U_\theta \mathbf{e}_\theta) = -\nabla \times (\Omega_{LO}(r) \mathbf{e}_z)$, with 
$\Omega_{LO}(r) = \Omega\exp(-r^2/a^2)$ the vorticity and $a=5$. 
\blu{The latter} velocity field is then compatible with periodic boundary conditions \cite{Dipierro:2012}. A plot of the initial distribution is shown in figure \ref{fig::DNS_TI} (top left).}
Note that the dynamics within the vortex core ($r \lesssim 0.3a$) is close to a solid body rotation. 
\pink{In the case of constant density, the 
Lamb-Oseen vortex does not satisfy the instability conditions of
 Leibovich \& Stewarson \citep{Leibovich:83}. Its linear stability 
has been proven numerically by \cite{fabre:2006}. }
Finally, the flow is initially disturbed with a random noise \blu{of amplitude
$O(10^{-3})$}.\\
The disturbance amplitudes $A_m$ of each mode $m$ are extracted from the Fourier expansion $\hat{\rho}_m(r, t)$ 
of the density field $\rho(r,\theta, t)$ interpolated from the Cartesian grid $(x, y)$, and are defined as:
\begin{equation}
A_m(t) = \sqrt{ \int_0^R |\hat{\rho}_m(r, t)|^2 r dr}.
\end{equation}

Figure \ref{fig::DNS_TI} shows a contour plot of the density field at times 
$t\approx 3.7, 5.5 \text{ and } 7.4$ 
for $Re=1000$ and $s=2$. \blu{At the earliest times}, many azimutal modes 
are growing (figure \ref{fig::DNS_TI} top right), while
the azimutal mode $m\approx17$ emerges a little later on (figure \ref{fig::DNS_TI} bottom left). This last mode is
the most unstable, that is predicted by the asymptotic expression $m_{max}$ (\ref{eq::M_most_unstable}). 

\begin{figure}
\centering
\includegraphics[width=0.5\linewidth]{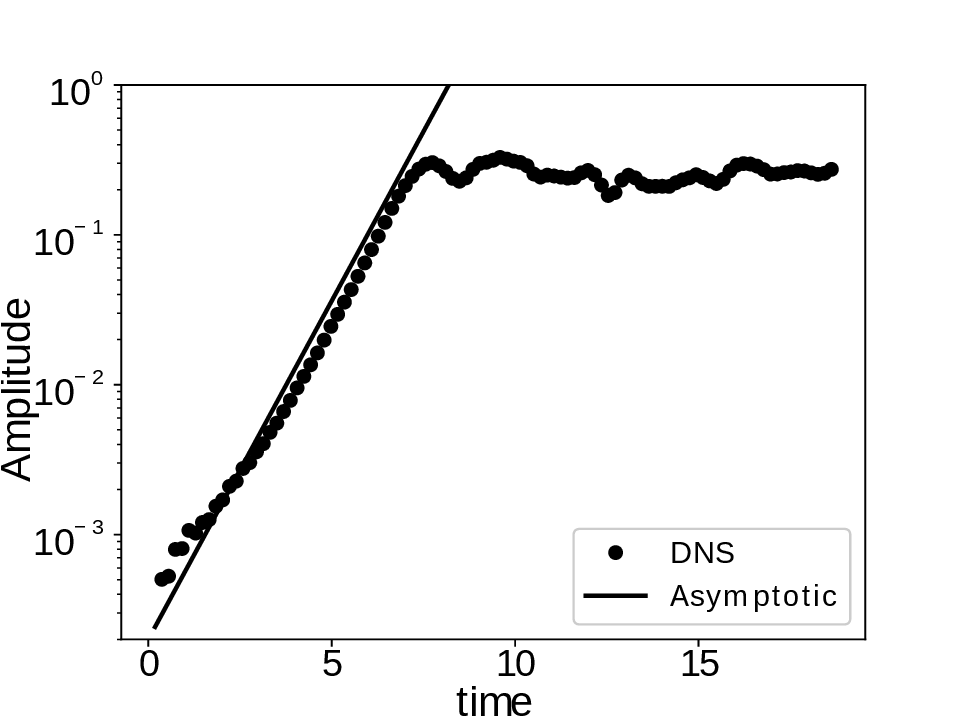}
\caption{Amplitude of the azimutal mode $m=17$ versus time, for $Re=1000$ and $s=2$ and the asymptotic prediction (\ref{eq::Omega_most_unstable},\ref{eq::M_most_unstable}). }
\label{fig::Amp_time}
\end{figure}

\begin{figure}
\centering
\includegraphics[width=0.42\linewidth]{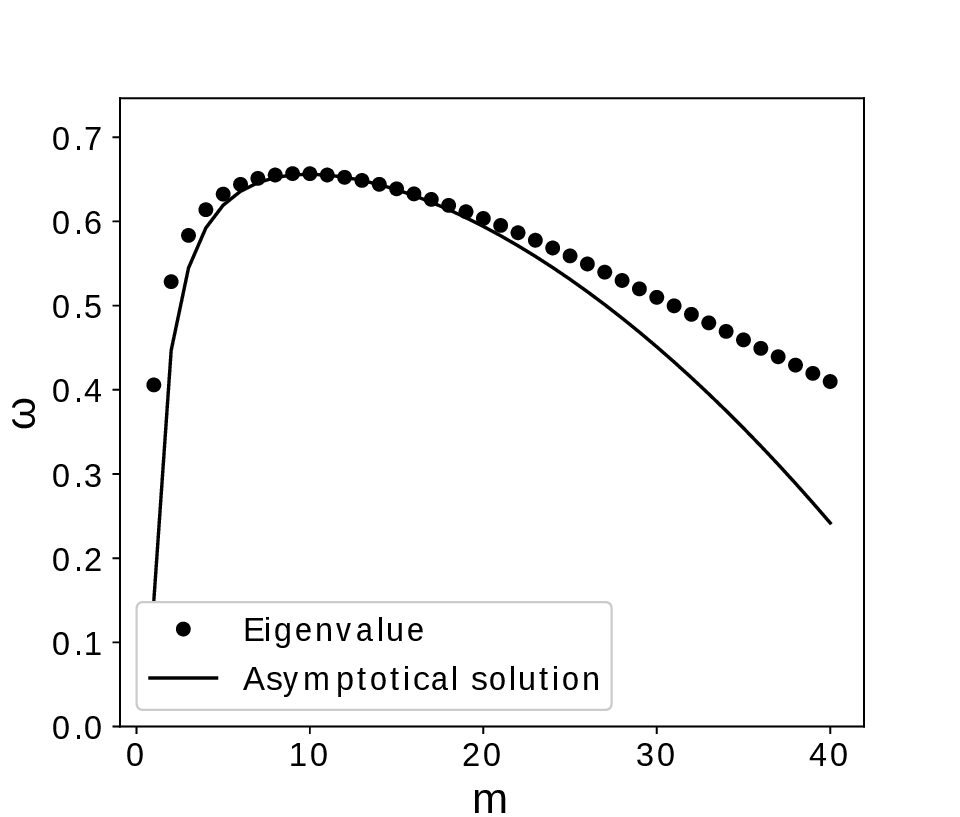}
\includegraphics[width=0.48\linewidth]{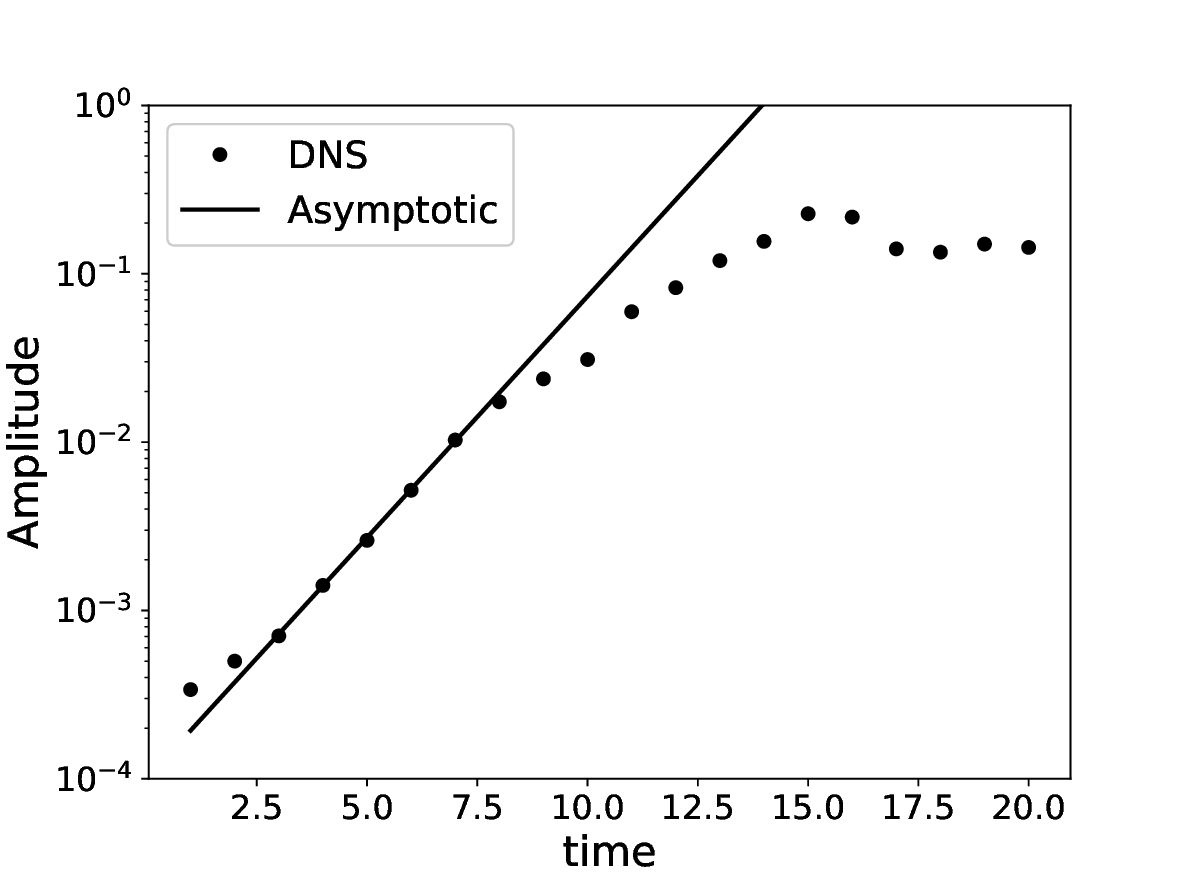}
\caption{Left: dispersion relation of base flow (\ref{eq::base2}). Right: Amplitude of the azimutal mode $m=10$ versus time, for $Re=1000$ and $s=2$ and the asymptotic prediction (\ref{eq::Omega_most_unstable}, \ref{eq::M_most_unstable}), for base flow (\ref{eq::base2}). }
\label{fig::Amp_time_exp}
\end{figure}

Figure \ref{fig::Amp_time} shows the amplitude of the $m=17$ mode versus time and the asymptotic prediction 
$\exp(-i\omega_{n, max}t)  $(see equations \ref{eq::Omega_most_unstable}, \ref{eq::M_most_unstable}); 
$Im(\omega_{n, max})\approx 1.04$ with the considered
parameters for this mode. After a transient time, one can see that the exponential
growth is well represented by the asymptotic theory. For $t\approx 7$, 
\blu{ figure \ref{fig::DNS_TI} (bottom right) shows that the instability
reaches a non-linear stage where modal interactions can no longer be
neglected.}
The study of these nonlinear interactions is left for future work.
A last simulation where $e/R=O(1)$ has been performed, with the basic flow: 
\begin{equation}
\rho_0(r) = 1 + (s-1)\exp(-r^2) \label{eq::base2}
\end{equation}
The comparison between direct numerical simulation and prediction of (\ref{eq::Omega_most_unstable}) is shown
on figure (\ref{fig::Amp_time_exp}). Once again, a very good agreement is found.

\section{Discussion and conclusion}
\pink{The paper presents a linear temporal stability analysis of an axisymmetric, variable-density, uniformly rotating flow
 evolving under the incompressibility constraint. }
\red{The problem solutions are validated by comparison to spectrally accurate numerical methods.}
\pink{In the inviscid case, which is considered first, the problem is governed by a second order differential equation.}
It is shown that the instability dynamics is \pink{characterized} by an analog to the Br\"unt-Va\"issalla
frequency $G(r)$ (with $G^2(r) = (-\Omega^2 r /\rho_0) d \rho_0/dr~$).
It is also shown that disturbances reach a maximum near the location where $G^2(r)$ admits an extremum. A non trivial 
dispersion relation is obtained. However, 
in the limit of large azimuthal wavenumbers, an approximation of eigenfrequency can be analytically obtained.
It can be seen that an instability
can occur if and only if $G^2>0$ \pink{(\textit{i.e.} if $d \rho_0/dr<0$)}. \tap{In that sense, it completes results given by
Scase and Hill \cite{scase:2018} restricted to the sharp initial density profile case and confined within a given
radial distance from the rotating center.}
 One deduces that the instability
is triggered when the local centrifugal acceleration (\red{$\mathbf{a_c}=\Omega^2 r \mathbf{e_r}$}) and the reduced density gradient
($\bar{\nabla \rho} = (\nabla \rho_0)/\rho_0$) are in opposite directions. This shows that the resulting
instability is a cylindrical Rayleigh--Taylor one, the disturbance growth rate being proportional to 
the maximum of $\sqrt{-\mathbf{a_c}\cdot\bar{\nabla \rho}}$. This is very similar to the classical planar
Rayleigh--Taylor growth rate $\sqrt{gkA_t}$ ($g,~k$ and $A_t$ being the gravitational acceleration, the
wavenumber and the Atwood number respectively). 
\pink{When comparing the expression of the eigenfrequencies at large $m$ for 
solid-body rotation to the asymptotic result of \cite{Dipierro:2010}
obtained assuming weakly-varying velocity profiles, it appears that 
the instability growth rates are the sames up to order $1/m$. The only 
difference being that a change is observed in the the real frequency of order $1/m$.}
\pink{In the inviscid case,  all modes are unstable and the instability  
growth rate is an increasing function of the azimuthal wavenumber 
approaching the asymptotic value $\sqrt{G^2}$ for large $m$.} \red{Hence,
no mode selection is observed.}
To correctly represent the mode selection, viscous effects have been studied in the limit of large
Reynolds numbers as well as large $m$. The resulting dynamics equation is very similar to the
inviscid one and viscosity appears as a purely stabilizing effect. Indeed, the real frequency is
not affected by the viscous drag while the growth rate varies as $-m^2/(\rho_0 Re)$. Once again, this is
remarkably similar to the planar Rayleigh--Taylor viscous correction proportional to $-k^2/(\rho_m Re)$
($\rho_m$ being an average density). Thus, the resulting instability
\blu{exhibits competitive mechanisms between the inviscid Rayleigh--Taylor
 instability and to viscous damping}.
The most unstable mode has a growth rate varying as $Re^{-1/3}$ while the corresponding azimuthal
wavenumber grows as $Re^{1/3}$. \red{The latter scaling is very similar to the
planar Rayleigh--Taylor instability for which the growth rate varies as
$Re^{-1/3}$, while the corresponding longitudinal wavenumber grows as
$Re^{2/3}$.} \\
\blu{Considering three dimensional disturbances for the same base flow (
\textit{i.e.} a columnar vortex), one notes that the dynamics equation
remains the same when considering the modified wave number 
$l^2 = m^2 + k^2r^2$, where $k$ the axial wave number.}
In particular, in the limit of large wavenumbers 
$l\rightarrow \infty$ and large Reynolds numbers, equation (\ref{eq::phivis}) remains invariant 
by substituting $m$ with $l$. Hence, the same instability characteristics
 are obtained. Such three-dimensional disturbances are \textit{helical} 
Rayleigh-Taylor instabilities. \tap{Such disturbances has been observed by Zeng \textit{et al.}  \cite{Zeng:2020}
for biphasic radially accelerated flows.}
Finally, these results have been validated using nonlinear direct numerical simulations \ble{where} the
instability is \blu{triggered by superimposing random noise of small 
amplitude onto the base flow}. The most unstable mode observed 
in these simulations
is in very good agreement with asymptotic theory. In the later 
times of the simulation, while nonlinearities appear, a strong modal
 interaction leads to the formation of mushroom-shaped disturbances, 
which are characteristics of a Rayleigh--Taylor instability. \tap{The resulting flow 
pattern is very similar to the one observed by Scase and Hill \cite{scase:2018}
in confined configuration.}
\section*{Acknowledgement}
The authors \ble{acknowledge the ``F\'ed\'eration Lyonnaise de Mod\'elisation et Sciences Num\'eriques}'' for
providing numerical facilities on the cluster P2CHPD.

\bibliographystyle{jfm}
\bibliography{rot_rhovar}

\end{document}